\documentclass[12pt, preprint]{aastex}

\usepackage{pdfsync}
\usepackage{amsmath}
\usepackage{natbib}

\DeclareMathSymbol{\varOmega}{\mathord}{letters}{"0A}
\DeclareMathSymbol{\varSigma}{\mathord}{letters}{"06}
\DeclareMathSymbol{\varPsi}{\mathord}{letters}{"09}

\newcommand{\Eq}[1]{Equation\,(\ref{#1})}
\newcommand{\Eqs}[2]{Equations (\ref{#1}) and~(\ref{#2})}

\begin{document}

%\slugcomment{Draft Modified \today}

\shorttitle{Superparticles}
\shortauthors{Youdin et al.}

\title{Superparticle Method for Simulating Collisions}
\author{David Nesvorn\'y$^1$, Andrew N.\ Youdin$^2$, Raphael Marschall$^1$, Derek C. Richardson$^3$}
\affil{(1) Department of Space Studies, Southwest Research Institute, Boulder, CO, USA}
\affil{(2) Steward Observatory, University of Arizona, Tucson, AZ, USA}
\affil{(3) Department of Astronomy, University of Maryland, College Park, MD, USA}

\begin{abstract}
For problems in astrophysics, planetary science and beyond, numerical simulations are often limited to 
simulating fewer particles than in the real system.  To model collisions, the simulated particles (aka 
superparticles) need to be inflated to represent a collectively large collisional cross section of real 
particles. Here we develop a superparticle-based method that replicates the kinetic energy loss during 
real-world collisions, implement it in an $N$-body code and test it. The tests provide interesting
insights into dynamics of self-gravitating collisional systems. They show how  
particle systems evolve over several free fall timescales to form central concentrations and 
equilibrated outer shells. The superparticle method can be extended to account for the accretional growth of objects 
during inelastic mergers.
\end{abstract}

\section{Introduction}
Numerical simulations are often limited in their ability to simulate statistically large particle systems.
For example, a protoplanetary disk simulation can account for $\sim10^8$-10$^9$ particles (and grid cells; e.g., 
Li et al. 2019), but the actual number of real-world elements (dust grains, boulders, etc.) is vastly larger. 
The simulated particles have more mass and represent a large number of real particles. We call this the 
superparticle method. The superparticle method has general applicability in many areas of science. Here we 
discuss it in the context of planet formation (see Youdin \& Kenyon (2013) for a review) mainly because our 
previous studies of the initial stages of planet formation directly motivated this work. 

During the earliest stages of planet formation, small grains condense in a protoplanetary nebula and grow 
to larger ice/dust aggregates by sticking to each other. The growth stalls near cm sizes because the 
electrostatic forces are not strong enough to hold large particles together. Also, large 
grains feel strong aerodynamic drag from the surrounding gas and drift toward the central star on timescales 
too short for significant growth to happen (e.g., Birnstiel et al. 2016, but see Michikoshi \& Kokubo 2016). 
It is therefore quite mysterious how planet 
formation bridges the gap between cm-size particle aggregates (hereafter {\it pebbles}) 
and 1--1000 km bodies ({\it planetesimals}; Chiang \& Youdin 2010, Johansen et al. 2014). 
A growing body of evidence now suggests that an aerodynamic interaction between pebbles and nebular gas (e.g., 
the streaming instability, Youdin \& Goodman 2005) can collect particles in large self-gravitating clouds that
then directly collapse into planetesimals.

The initial stages of particle concentrations in a gas nebula can be studied with specialized hydrocodes (e.g., 
{\tt ATHENA }with the particle module of Bai \& Stone 2010). Simulations show that the streaming instability 
(hereafter SI) should be particularly important \citep{yj07, jy07, jym09, nesvorny19, li19}. The instability occurs 
because an initially small over-density of pebbles accelerates 
the gas. This perturbation launches a wave that amplifies pebble density as it oscillates. Strong pebble clumping 
eventually triggers gravitational collapse into planetesimals. Modern hydrocode simulations of the SI, however, 
do not have adequate spatial resolution to follow the gravitational collapse to completion. Moreover, once the 
pebble density within a collapsing cloud exceeds the gas density, aerodynamic effects of gas cease to be important, 
and detailed hydrodynamic calculations are no longer required (Nesvorn\'y, Youdin \& Richardson 2010; hereafter NYR10). 
Instead, one has to realistically model pebble 
collisions that damp random speeds and stimulate growth (Wahlberg Jansson \& Johansen 2014). 

In our previous work (NYR10), we studied gravitational collapse 
with a cosmological $N$-body code known as {\tt PKDGRAV} (Stadel 2001). {\tt PKDGRAV} is a scalable, parallel tree 
code that is the fastest code available to us for these type of calculations. A unique feature of {\tt PKDGRAV} 
is the ability to rapidly detect and realistically treat collisions between particles (Richardson et al. 2000). 
In NYR10, individual {\tt PKDGRAV} particles were artificially inflated to mimic a very large 
collisional cross section of pebbles. We simply scaled up the radius of {\tt PKDGRAV} particles by a multiplication 
factor that was the same for all particles and unchanging with time.  This is not ideal for several 
different reasons. Crucially, the method with fixed inflation factor does not correctly account for 
the kinetic energy loss during inelastic collisions. 

Here we extend the superparticle method to be able to more realistically model planetesimal formation (Section 3). 
The new method is designed to reproduce the kinetic energy loss. It will be useful in many areas of science where 
particle collisions and energy dissipation are important. The method is tested for self-gravitating collisional 
systems in Section 4. To provide some background for these tests, we discuss several relevant timescales in 
Section 2. Section 5 summarizes our findings.

\section{Collision and collapse timescales}

\subsection{Collisions}
Here we define the collisional timescale, $t_{\rm coll}$, as the timescale on which any particle 
in a cloud would, on average, have one collision with another particle. The collisional timescale 
is then $t_{\rm coll}=1/\eta \sigma v$, where $\eta$ is the number density of particles, 
$\sigma=\pi r^2$ is the particle cross-section, $r$ is the particle radius, and $v$ is the relative speed. 
Adopting the virial speed $v=(G M / R)^{1/2}$, where $G$ is the gravitational constant and $M$ and $R$ are the 
mass and radius of the cloud, we obtain 
\begin{equation}
t_{\rm coll} = {4 \over 3 n r^2 \sqrt{G M}} R^{7/2}\ ,
\label{tau}
\end{equation}
where $n$ is the total number of cloud particles. The collisional timescale is therefore a steep 
function of cloud's size. We assume that the cloud radius is some fraction of the Hill 
radius, $R=f_{\rm H} R_{\rm H}$, where
\begin{equation}
R_{\rm H} = a \left( {M \over  3 M_\odot } \right)^{1/3}\ ,
\end{equation}
$a$ is the orbital radius, and $M_\odot \simeq 2\times10^{33}$ g is the solar mass. 
After substituting for $R_{\rm H}$ and $n=(R_{\rm eq}/r)^3$, where $R_{\rm eq}$ is the equivalent radius of 
a sphere with mass $M$ and density 1 g cm$^{-3}$, we obtain
\begin{equation}
t_{\rm coll} \simeq 23\, {\rm yr}\ \left( {r \over 1\, {\rm cm}} \right)
\left( {50\ {\rm km} \over R_{\rm eq}} \right)
\left( {f_{\rm H} \over 0.5}\, { a \over 45\, {\rm au}} \right)^{7/2}\ .
\label{tau2}
\end{equation}
For comparison, the orbital period at 45 au is $P\simeq300$ yr. This illustrates the 
importance of particle collisions. Collisions will act to damp particle speeds and stimulate 
growth.

\subsection{Virial collapse}

In virial equilibrium, $T=-U/2$, where $T=\sum_i m_i v_i^2 / 2$ and $U=-\sum_{i,j} 
Gm_im_j/r_{ij}$, with $r_{ij}=|\mathbf{r}_i-\mathbf{r}_j|$, are the total kinetic and 
potential energies of particles with masses $m_i$, positions $\mathbf{r}_i$ and 
speeds $v_i$. If a cloud is in virial equilibrium, it cannot collapse unless some 
dissipative process, such as inelastic collisions between particles, reduces the kinetic 
energy. In a single collision of two pebbles with masses $m_i'$ and $m_j'$, the 
change in energy is
\begin{equation}
\delta E = -{1 \over 4} \mu \Delta v^2 (1-C^2_{\rm R})\ , 
\label{de}
\end{equation}
where $\mu=m_i' m_j'/(m_i'+m_j')$, $\Delta v$ is the collision speed and $C_{\rm R}$ is the 
coefficient of restitution. The above
equation takes into account that collisions are not necessarily head-on, which reduces 
the average dissipated energy by a factor of 1/2.

Assuming that all particles have the same mass and $\Delta v = \sqrt{2}\, v$, 
where $v=(G M / R)^{1/2}$ is the virial speed, the total energy lost in the time interval 
$\Delta t$ is 
\begin{equation}
\Delta E = \delta E \times n {\Delta t \over t_{\rm coll}}    \ , 
\label{dde}
\end{equation}   
where $n$ is the number of particles and $t_{\rm coll}$ is given in Eq. (\ref{tau}).
This leads to a differential equation for the total energy of the cloud, $E$, or 
equivalently, for the cloud radius. Wahlberg Jansson \& Johansen (2014) showed that
\begin{equation}
E = E_0 \left( 1 - {t \over t_{\rm vir}} \right )^{-2/7}\ , 
\label{energy}
\end{equation}
where $t$ is time and $E_0$ is the initial energy. This assumes that frequent collisions 
instantaneously virialize the cloud. The virial collapse timescale $t_{\rm vir}$ is given by 
\begin{equation}
t_{\rm vir}= 0.26\, t_{\rm coll} (1-C^2_{\rm R})^{-1}  
\end{equation}
with $t_{\rm coll}$ in Eq. (\ref{tau}). The virial collapse timescale is thus
roughly four times shorter for $C_{\rm R}=0$ than the collisional timescale for $t=0$.
This is a consequence of the steep dependence of $t_{\rm coll}$ on $R$. In the virial 
collapse, the cloud radius $R$ is related to Eq. (\ref{energy}) via
\begin{equation}
R={3 G M^2 \over 10 E}\ .
\end{equation} 

\subsection{Free fall collapse}\label{sec:freefall}
The above analysis is valid when the collapse is ``hot'', i.e., when particle speeds 
are virial and remain virial during the whole collapse. An alternative to this is the 
``cold'' collapse with sub-virial particle speeds. In this case, the collapse timescale 
is related to the free fall timescale 
\begin{equation}
t_{\rm free}=\left({3 \pi \over  32 G \rho_{\rm cloud}}\right)^{1/2} \ , 
\label{ff}
\end{equation} 
where $\rho_{\rm cloud}=3 M/ 4 \pi R^3$ is the cloud's mass density. This assumes that 
particles have negligible initial velocities and collisions are ignored. If the mass is 
uniformly distributed within a spherical volume of radius $R$, the collapse is self-similar 
and all radial shells shrink on the timescale given by Eq. (\ref{ff}). If the mass is initially 
concentrated toward the cloud's center, the inner shells will infall faster than the 
outer shells.  

Substituting $\rho_{\rm cloud}$ and $R=f_{\rm H} R_{\rm H}$ into Eq. (\ref{ff}), we obtain
\begin{equation}
t_{\rm free} = 11\,{\rm yr}\ \left( {f_{\rm H} \over 0.5} \, {a \over 45\, {\rm au}} \right)^{3/2}\ ,
\label{ff2}
\end{equation}
which is independent of $M$ (or $R_{\rm eq}$), because more massive clouds have larger Hill 
radii and these dependencies cancel out in $\rho_{\rm cloud}$. Thus, in our fiducial case 
with $f_{\rm H}=0.5$, $a=45$~au, $r=1$ cm, $R_{\rm eq}=50$ km and $C_{\rm R}=0$, we obtain 
$t_{\rm vir} < t_{\rm free}$, which shows that collisions damp the random velocities 
faster than the cloud can collapse. The collapse will thus happen on the free 
fall timescale. Given the different scaling of Eqs. (\ref{tau2}) and (\ref{ff2}) with 
$R_{\rm eq}$, only small clouds with $R_{\rm eq} < 25$ km can, at least initially, contract on the 
$t_{\rm vir}$ timescale. As $t_{\rm coll}$ drops faster during the collapse than $t_{\rm free}$, 
however, all clouds will eventually free fall.\footnote{The considerations ignore the accretional 
growth of particles during collisions, which acts to increase $r$, $t_{\rm coll}$ and $t_{\rm vir}$.
Particle fragmentation would have an opposite effect (Wahlberg Jansson et al. 2017).}

\section{Superparticle method}
Two particle systems are considered here: one that consists of real particles (RPs) and another 
one that consists of simulated superparticles (SPs). Inelastic collisions between particles happen in both systems. 
They can result in inelastic bounces or mergers of particles, and the loss of kinetic energy. We assume that the number of 
SPs is much smaller than the number of RPs, and ask how to best deal with collisions in the SP system such as it 
statistically reproduces the behavior (i.e., energy loss, particle growth) of the RP system. 

\subsection{Size distributions}
We denote the number distributions of SPs as $dN$ and RPs
% DN changed this from true particle because TPs looks like test particles
as $dN'$. In terms of the RP radius $r$ (no prime here) we may have a size distribution 
$dN'/dr \propto r^{-q}$. The number distribution of SPs can (i.e., might need to) be different, e.g. 
$dN/dr \propto r^{-Q}$.\footnote{Note that $dN/dr$ is not the same as the distribution $dN/dR$, where 
$R$ denotes the SP size; the relationship between RP's $r$ and its SP's $R$ is not specified at 
this point.} For instance, $Q = 1$ would be a log-uniform distribution that might be a good choice 
to numerically sample. 

We define $n(r)$ as the number of RPs replaced by a single SP, 
or (equivalently) the mass ratio of a SP and its RP. The mass in each bin is the same in both systems, 
$dM = m(r) dN = m'(r) dN'  = n(r) m'(r)dN $, where $m'(r)$ is the mass of a RP.  Thus 
\begin{align}\label{eq:n}
 n(r) &=  dN'/dN = m/m'
\end{align} 
and for the example of power-law distributions $n(r) \propto r^{Q-q}$ and $n(r) \propto r^{1-q}$ for log-uniform 
SPs.  With typical $q \approx 3$--$4$, we see that the sampling rate $n(r)$ should be much larger for 
smaller RPs.  Indeed for constant density RPs with $m' \propto r^3$, our log-uniform example means that SPs 
for smaller $r$ will be {\it more} massive for $q > 4$ (or $q > 3 + Q$ for other SP power-laws). We 
will refer to this pathological case as SP mass inversion.

More generally, for non-constant $n(r)$, the mass ratios of SPs will not be the same as the RPs they model.  
This may have some undesirable properties for gravitational and collisional dynamics.  In particular, both 
gravitational scattering and (inelastic) collisions tend towards equipartition, i.e., equal kinetic energies 
for all species.  If SPs do not have the same $n$, then we risk forcing equipartition of the artificial 
SP masses.  Since varying $n(r)$ may be difficult to avoid, the question is whether this tendency to 
equipartition is significant, or whether it is over-ridden by other physics (e.g., the fact that collisions 
are inelastic).  Moreover the timescale for equipartition could be long compared to timescales of 
interest (e.g., for gravitational collapse).  
%We test this in Section 4.

\subsection{Inelastic collisions}\label{sec:coll}
Intuitively the system of SPs must have larger collisional radii than RPs to account for the reduced number.   
Specifically, we want the energy loss rate from collisions per unit volume, $\dot{E}$,  to be the same in the 
RP and SP systems.  For collisions between species $i$ and $j$ the differential loss rate is 
\begin{align}\label{eq:edot}
d\dot{E}_{ij} = \frac{1}{4} dN (r_i) dN (r_j) \sigma_{ij}  v_{ij}^3 \mu_{ij} f(C_{\rm R})
\end{align}
for cross section $\sigma_{ij}$, relative speed $ v_{ij}$, reduced mass $\mu_{ij}=m_i m_j/(m_i + m_j)$ and 
coefficient of restitution $C_{\rm R}$. Averaged over impact angles, $f(C_{\rm R}) = (1-C_{\rm R}^2)/2$ (Section 2.2).  
The total $\dot{E}$ is obtained by a double integral over the size distributions for species $i$ and $j$.  In more 
detail, the above equation is already averaged over a velocity distribution and impact angles, and the $dN$'s are 
now per unit volume.  The 1/4 in Eq. (\ref{eq:edot}) represents 1/2 from kinetic energy and 1/2 to avoid double counting of species 
pairs.

We now equate our (unprimed) SP system with the (primed) RP system, $d\dot{E}_{i,j} = d\dot{E}_{i,j}'$. We assume 
$ v_{ij} =  v_{ij}'$ and $f(C_{\rm R}) = f'(C_{\rm R})$, i.e., that the SPs have a similar velocity distribution 
to the RPs and there is a similar distribution of impact angles.  From \Eqs{eq:n}{eq:edot} we get the desired 
result:
\begin{align}\label{eq:sigij}
\sigma_{ij} &= \sigma'_{ij} \frac{m_i +  m_j}{m_i' + m_j'} = \sigma'_{ij} \frac{n_i m_i' + n_j m_j'}{m_i' + m_j'}\ ,
\end{align} 
i.e., the cross section to mass ratio is the same for SP and RP collisions. 

If SPs $i$ and $j$ represent the same number of RPs, $n_i=n_j$, then $\sigma = \sigma' n$, 
i.e., the effective SP radii are $R_i = \sqrt{n} r_i$ and $R_j = \sqrt{n} r_j$. For strongly unequal true masses $m_i' \gg m_j'$, 
\begin{align}
\sigma_{ij}  &\xrightarrow{m_i' \gg m_j'} \frac{\pi r_i^2}{m_i'} (m_i +  m_j)\ .
\end{align}   
If the SP masses are not inverted, such that $ m_i > m_j$, we again get $R_i \approx \sqrt{m_i/m'_i} r_i = \sqrt{n_i} r_i$.  
Moreover, if we define (now relaxing the assumption of strongly unequal masses)
\begin{align}
\sigma_{ij} &\equiv \pi (R_i + R_j)^2 \equiv \pi (\sqrt{n_i} r_i +  R_j)^2\ ,
\end{align} 
then $R_j > 0$ by application of  \Eq{eq:sigij}.  This result shows that even the naive estimate of 
$R_i \approx  \sqrt{n_i} r_i$ is ``safe" in that it won't require negative radii for the collisional partner, i.e.\ the other SP.  

\subsection{Mergers}

If SPs merge, this represents a merger of some large number of RPs.  If SPs $i$ and $j$ merge to a new 
SP $k$, the masses simply add $m_k = m_i + m_j$.  However there is some choice as to how the 
underlying true particle masses combine.  We would like to exploit this choice so that SPs gradually 
approach RPs as coagulation proceeds.  The final collapse outcome will then have more realistic densities 
and a more meaningful final multiplicity (single, multiple, etc.).  The obvious choice of simply 
merging the true masses, as $m_k' = m_i' + m_j'$,  doesn't allow this transition, as we now explain 
in more detail. 

Our desired transition  requires a gradual decrease in the sampling rates $n_k$.  The obvious merging 
strategy clearly doesn't have this property.  If all SPs merge they have a total mass $m_{\rm tot} = 
\sum_i m_i = \sum_i n_i m_i'$, while the simple sum of RP masses is $m'_{\rm tot} =  \sum_i m_i'$ (where 
$i$ is now a sum over each SP and not each size bin).  Clearly, the sampling rate of this final object 
$n_{\rm tot} =  \sum_i n_i m'_i/\sum_i m'_i$ will be large if the initial $n_i$ are large, as this final 
$n_{\rm tot}$ is just a mass-weighted average of the initial $n_i$.  Again, a decrease of the sampling 
rate is needed to approach a true system.

During a merger event, we thus choose to reduce the merged sampling rate $n_{k, \text{merged}}$ and thus 
increase the RP masses (since the SP masses must be conserved).   From the merger of $m_k = m_i + m_j$,  
the  post-merger sampling rate must obey
\begin{align}
n_{k, \text{merged}} &< \frac{m_k}{m'_i + m'_j}  \equiv n_{k, 0}\, , 
\end{align} 
where $ n_{k, 0}$ would arise from the obvious summation of RP masses. For instance, in Section 4.3, we 
consider $n_{k, \text{merged}}=   \max(n_{k, 0}(1 - m_j/m_k), 1)$ for $m_i > m_j$.  (Note that this correction uses 
the SP masses, but we also have $m'_i > m'_j$ for the RP masses, provided we stick to the non-inverted 
samplings.) 

With different merger prescriptions, we should consider how the mass growth timescales of SP and RP 
systems are related in theory.  Recall that we are considering an SP cross section that correctly 
models energy dissipation but not necessarily the mass growth.  To explore this issue, we consider 
the mass growth rate of a given SP in bin $i$ due to the SPs in bin $j$ : 
\begin{align}
\frac{\dot{m}_{i,j}}{m_i} &= \frac{m_j}{m_i} dN_j \sigma_{ij} v_{ij}
\end{align} 
and the corresponding RP growth rate (for simple addition of RP masses, i.e. before any rescaling) :
\begin{align}\label{eq:RPgrowth}
\frac{\dot{m}'_{i,j}}{m'_i} &= \frac{m'_j}{m'_i}  dN'_j \sigma'_{ij} v'_{ij} = \frac{m_j}{m'_i}  
dN_j \sigma_{ij} v_{ij} \frac{m'_i + m'_j}{m_i + m_j} = \frac{\dot{m}_{i,j}}{m_i} 
\left[ \frac{1 + m'_j/m'_i }{1 +  m_j/m_i} \right]\, .
\end{align} 
Above we use our previous result for conservation of mass per bin ($m_j dN_j = m'_j dN'_j$), and again use 
the cross section relation of \Eq{eq:sigij} assuming accurate modeling of velocities $v_{ij} = v'_{ij}$.  
\Eq{eq:RPgrowth} shows that the mass growth timescales for the RP and SP systems 
will agree in the special case where the term in square brackets is unity.  This special condition 
occurs  (1) for uniform sampling, i.e.\ $n_i = n_j$, which again is often overly restrictive 
and (2) in the limit of very small accreted masses $j$, both for the SP and RP masses.
Aside from these special conditions, mass growth rates will not agree exactly.  We thus confirm that a cost of 
non-uniform SP sampling is that energy dissipation and mass growth can't both be reproduced exactly.  

In what direction should the inexact growth rates go?  Consider $m_i' > m_j'$ and the standard case 
$n_j > n_i$ for more efficient sampling.  Then \Eq{eq:RPgrowth} gives $\dot{m}_{i,j}/m_i > \dot{m'}_{i,j}/m'_i$.  
In words, the large SPs will grow at a statistically larger rate than they ``should" if they were 
accurately modeling the RP system.\footnote{Our post-merger reduction of the sampling rates $n_k$ doesn't affect 
the SP masses, but rather the RP masses.  With decreasing $n_k$ the SP growth rates approach those of the RPs, 
but since the RPs have been modified, this is not the same result as the (too difficult) simulation of RPs from the outset.}
In practice, differences in the SP vs. RP relative velocities will also affect  growth rates.  Thus we must compare 
RP and SP growth rates numerically to asses these issues quantitatively.  

\subsection{{\tt PKDGRAV} implementation}
The method described above was implemented in the hard-sphere flavor of {\tt PKDGRAV} (Richardson et al. 2000). 
To detect collisions, {\tt PKDGRAV} sorts neighbors of each particle by distance. The particle and each neighbor are then 
tested for a collision during the next timestep ${\rm d}T$. The condition is simply $d_{ij}(t)\leq 
R_i+R_j$, where $d_{ij}(t)$ is the distance of $i$ and $j$ particles at time $0<t<dT$, and $R_i+R_j$ is the sum 
of particle radii. We changed the collision condition in {\tt PKDGRAV} such that $R_i+R_j=\sqrt{\sigma_{ij}/\pi}$, 
where $\sigma_{ij}$ is given by Eq. (\ref{eq:sigij}). When a collision is detected, {\tt PKDGRAV} decides, based on 
user-defined parameters, collision speed, etc., whether it results in a merger or bounce. 

If a merger is applied, we use the method described in Section 3.3 to generate a new SP. {\tt PKDGRAV} 
then computes the velocity of the new SP from the linear momentum conservation. Particle bounces in {\tt PKDGRAV} 
are controlled by the normal coefficient of restitution, $0 \leq C_{\rm R} \leq 1$, where $C_{\rm R}=0$ corresponds to a 
fully inelastic collision and $C_{\rm R}=1$ to an ideally elastic bounce (i.e., no energy dissipation). 
The code computes post-collision velocities of SP $i$ and $j$ and moves to considering a new SP pair. 
{\tt PKDGRAV} is also capable of producing particle aggregates, where interacting particles become rigidly 
locked as a group, but we do not use this option here.    

Specifically, we implemented the following changes in {\tt PKDGRAV}:
\begin{enumerate}
\item We modified the structure of the input/output files such that $n_i$ and $r_i$ of each SP is given 
in the last two columns. New variables, {\tt fNpebble} and {\tt fRpebble} were included in the {\tt PKD} (particle data) 
and {\tt COLLIDER} (collision data) structures. The new variables are passed between these structures in 
{\tt pkdGetColliderInfo} and {\tt pkdPutColliderInfo} (functions to retrieve and store collision data). 
{\tt SSDATA\_SIZE} (solar system data size) in {\tt ssio.h} was increased to make space for new variables.
The function {\tt pkdReadSS} now reads these variables from input and passes it to the {\tt PKD} structure. 
\item We changed the function {\tt CheckForCollision} (function that tests particle's neighbors for collision
in the current timestep) in {\tt smoothfcn.c} such that the sum of particle radii 
is ${\tt sr}=\sqrt{\sigma_{ij}/\pi}$, where $\sigma_{ij}$ is given in Eq. (\ref{eq:sigij}). 
We also modified option {\tt OverlapAdjPos} (a method to deal with overlapping particles by separating them along 
their line of centers) in {\tt pkdDoCollision}, included in {\tt collisions.c}, such that the displacement is 
aligned with the definition of {\tt sr} in CheckForCollision.
\item We implemented the merger algorithm from Sect. 3.3 in function {\tt pkdMerge}, included in {\tt collisions.c}. 
This was done in two steps to account for the merger of $i$ and $j$ SPs, and to reduce the number of RPs. In 
{\tt pkdBounce}, the individual particle radii were replaced by $R_i=r_i \sqrt{(n_i m_i' + n_j m_j')/(m_i' + m_j')}$ and 
$R_j=r_j \sqrt{(n_i m_i' + n_j m_j')/(m_i' + m_j')}$. This assures that the linear momentum conservation part 
in {\tt pkdBounce} operates with the right quantities. The SP bounces in {\tt PKDGRAV} are treated as bounces of 
particles with inflated radii.
\item The expression for the escape velocity of colliding SPs in {\tt pkdDoCollision} (function that executes collisions) 
was changed such that $v_{\rm esc}^2=2(m_i+m_j)/\sqrt{\sigma_{ij}/\pi}$ with $\sigma_{ij}$ from Eq. (\ref{eq:sigij}). 
{\tt PKDGRAV} uses the escape velocity and the parameter {\tt dMergeLimit} to decide whether the SPs should be merged.  
\end{enumerate}
The modified PKDGRAV code works with the {\tt pthread} parallelization and we typically used 5, 10 or 28 Broadwell cores 
in the tests described in Sect. 4.

\section{Tests}

The highest-resolution SI simulations published to date produce self-gravitating clouds with up to $\sim$$10^6$ 
SPs and cloud masses corresponding to solid planetesimals from tens to hundreds kilometers accross. 
To focus on a specific scenario, consider a pebble cloud with $N=10^6$ SPs and total mass $M\simeq5\times10^{20}$ g 
(corresponding to a solid planetesimal with $R_{\rm eq}=50$ km and density $\rho =1$ g cm$^{-3}$), and assume that 
all pebbles have the same radius $r \simeq 10$ cm (this is merely a convenient choice for the tests described below; 
pebbles in the outer solar system are expected to be at least $\sim$10 times smaller, Birnstiel et al. 2016).
The total number of pebbles is then 
%AY: I removed n here since it has another meaning
$3 M / 4 \pi r^3 \rho \sim 10^{17}$. We thus see that the number of pebbles exceeds the number of 
SPs by $\sim$11 orders of magnitude. From Section \ref{sec:coll}, the collisional radius of each SP 
would need to be $R_{\rm SP} \approx 10 \sqrt{10^{11}}\textrm{ cm} \simeq 32 \textrm{ km}$ to have 
a correct rate of energy dissipation. 

The tests described below were designed to mimic the situation in a self-gravitating, collisional cloud
of particles. Since we are unable to simulate the actual number of pebbles, we consider a reduced system with 
$10^6$ RPs distributed in a spherical volume with $f_{\rm H}=0.1$, $a=45$ au and $R_{\rm eq}=50$ km. If RPs 
were given a solid density 1 g cm$^{-3}$, then $r \simeq 0.5$ km and the collisional timescale would 
be excessively long ($t_{\rm coll} \simeq 10^3$ yr according to Eq. (\ref{tau})). We therefore scale up the  
radii to have $r=32$ km. This is done to reproduce the collisional timescale in the $10^{17}$ pebble 
cloud case discussed above. With $10^6$ particles and $r=32$ km, we have $t_{\rm coll} \simeq 0.8$ yr, whereas 
$t_{\rm free} \simeq 1$ yr from Eq. (\ref{ff2}). The collision and free fall timescales are therefore initially 
similar.\footnote{The effects of aerodynamic gas drag are not considered here. See Nesvorn\'y et al. (2010)
for a justification of this assumption in more realistic pebble collapse simulations.}

To test the SP method, we keep $f_{\rm H}=0.1$, $a=45$ au and $R_{\rm eq}=50$ km, and represent 
the particle cloud by $10^2$--$10^5$ SPs. This means that each SP initially represents $10$--$10^4$ RPs. 
The results of these tests were compared to the fiducial case with $10^6$ RPs. This is a good way to test the 
validity of the SP method because we know what the ground truth is (as given by the $10^6$ RP simulation).
Specifically, we considered the energy dissipation, radial and velocity distributions, angular momentum 
conservation, etc. If mergers were applied, we also compared the size distributions.   

Initially, particles were uniformly distributed within a sphere of radius $f_{\rm H}R_{\rm H}\simeq30,000$~km 
and were given slightly sub-virial speeds (see below). We also considered other initial conditions such the isothermal 
radial profile from Binney and Tremaine (1987) or locally virial conditions, where $T=-U/2$ was imposed at each 
radius. The particle density of the isothermal radial profile has a singularity near the cloud's center. This is 
not convenient because simulated particles become crowded in the center and the code slows down as it must deal 
with a vast number of collisions. The locally virial cloud is not in an equilibrium and evolves toward becoming 
globally isothermal. The character of this evolution is similar, except for the initial stage, to that reported for
our standard (i.e., uniform and sub-virial) setup below.    

\subsection{Elastic collisions}
We first consider a case with $C_{\rm R}=1$ (fully elastic collisions) to verify the energy conservation in {\tt PKDGRAV}. 
Particles were given random speeds $v_{\rm rand}=0.8$ m s$^{-1}$, which is slightly lower than the virial speed 
($v \simeq 1.1$ m  s$^{-1}$). This corresponded to $T/|U|=0.47$ initially. The slightly sub-virial conditions 
were used here to encourage collapse and mimic conditions that might exist in a self-gravitationg clump of pebbles 
just after it formed by the SI. The {\tt PKDGRAV} timestep was set to ${\rm d}T=0.001$ yr and the whole 
simulation covered 10 yr, which is about ten free fall timescales. 

We tested cases with different {\tt PKDGRAV} tree opening angles ({\tt dTheta}=0.2, 0.5 and 1) and found that the 
fractional change of the total energy was $2\times10^{-5}$ for {\tt dTheta}$=0.2$ and 0.5
and $10^{-4}$ for {\tt dTheta}$=1$. This result is expected since for smaller opening angles, cells must be further 
away to be treated in the multipole approximation, and the force calculation is more accurate. The simulations with 
$10^6$ RPs were completed in 7 (for {\tt dTheta}$=0.2$), 3 ({\tt dTheta}$=0.5$) and 2 ({\tt dTheta}$=1$) hours of processing 
time on 28 Broadwell cores. Decreasing the timestep to ${\rm d}T=0.0001$~yr resulted in only a modest improvement of the energy 
conservation. We thus used ${\rm d}T=0.001$ yr and {\tt dTheta}$=0.5$ in the following tests.
The system remained near the virial equilibrium with $T/|U|$ monotonically increasing from the initial 0.47 to 
the final 0.51. 

\begin{figure}
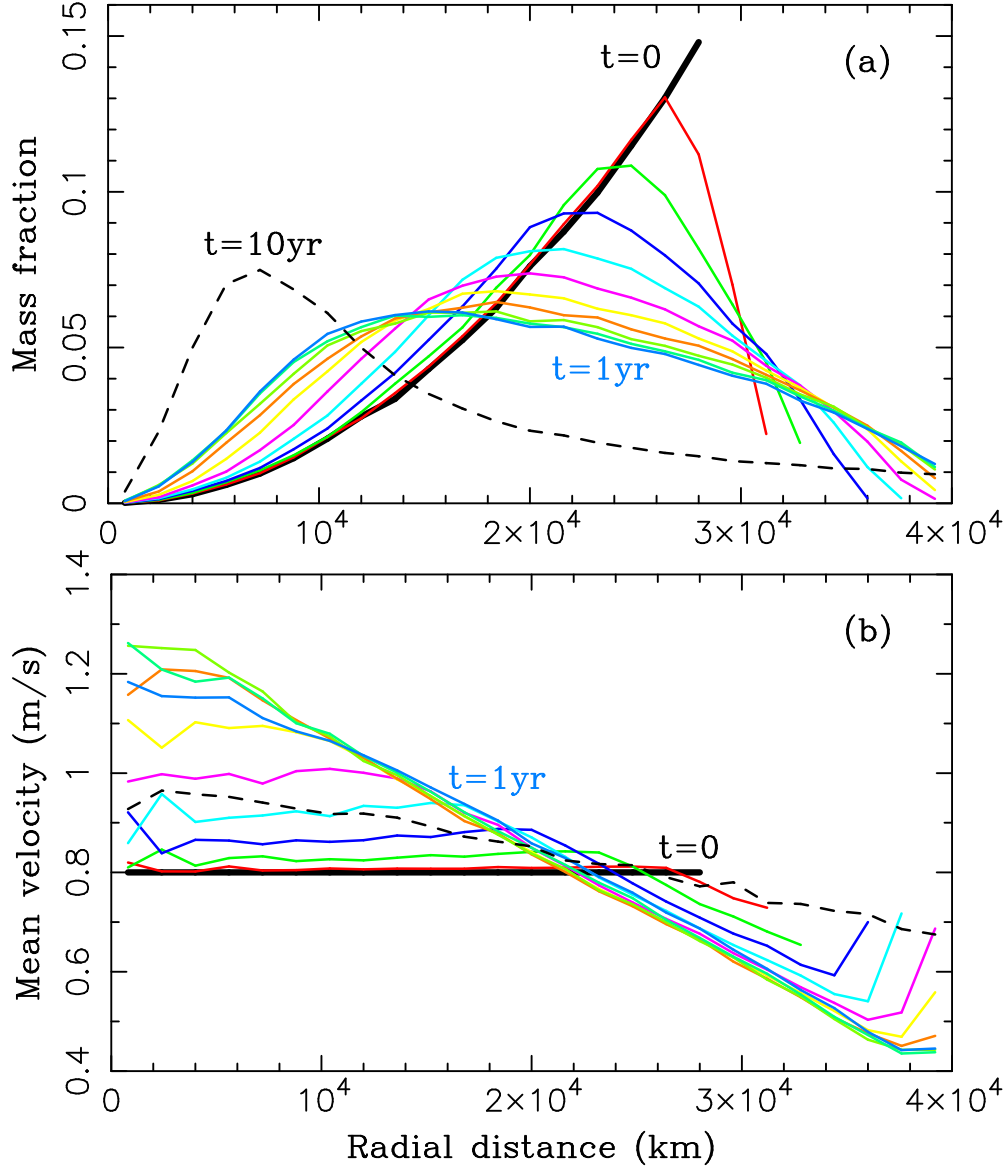

\epsscale{0.8}
\plotone{fig1a.eps}
\plotone{fig1b.eps}
\caption{The mass (a) and mean velocity (b) profiles of a self-gravitating collisional cloud. Initially, 
$10^6$ particles with $r\simeq35$ km were homogeneously distributed in a spherical volume with $f_{\rm H}=0.1$, 
$a=45$ au and $v_{\rm rand}=0.8$ m s$^{-1}$. Here we used $C_{\rm R}=1$ (fully elastic case). The solid and dashed 
lines show the initial ($t=0$) and final ($t=10$ yr) profiles. The colored lines follow profiles at 0.1 yr 
increments from $t=0$ to 1 yr.}
\label{profile}
\end{figure}

Figure \ref{profile}a shows the radial mass distribution of the cloud. We divided the integration domain, represented 
by a sphere of radius 40,000 km, into 25 concentric shells of equal radial width, ${\rm d}{\cal R}$, and plot the mass in each 
shell as a fraction of the total  mass ($M\simeq5\times10^{20}$~g). Mathematically, the plotted quantity is 
${\rm d} M({\cal R})/M = 4 \pi \rho {\cal R}^2 {\rm d}{\cal R}/M$,  where $\rho$ is the mass density at radius ${\cal R}$. % and ${\rm d}{\cal R}$ is the radial bin size. 
For the  singular isothermal profile with $\rho \propto {\cal R}^{-2}$ (Binney \& Tremaine 1987),  
${\rm d} M({\cal R})=$ const. Thus our mass profiles highlight differences relative to $\rho \propto {\cal R}^{-2}$.

By design most mass is initially contained in the radial shells near the outer edge of the cloud. Over one free 
fall timescale the cloud contracts and the radial mass distribution has a maximum near 15,000 km. This trend 
continues during the subsequent evolution until eventually, by the end of the simulation, the maximum 
mass fraction is near 7,000 km. This is a by-product of particle inflation. The total volume of 
the inflated particles is equivalent to a sphere of radius $\simeq$3,500 km. Thus, as the cloud 
contracts, the center becomes crowded and the mass of the inner shells cannot increase beyond 
certain limits. The final profile in the outer regions approaches a theoretical profile of the
isothermal sphere, where our mass distribution is expected to be independent of the radial distance 
(Binney \& Tremaine 1987; their Eq. 4-123). 

\begin{figure}
\epsscale{0.8}
\plotone{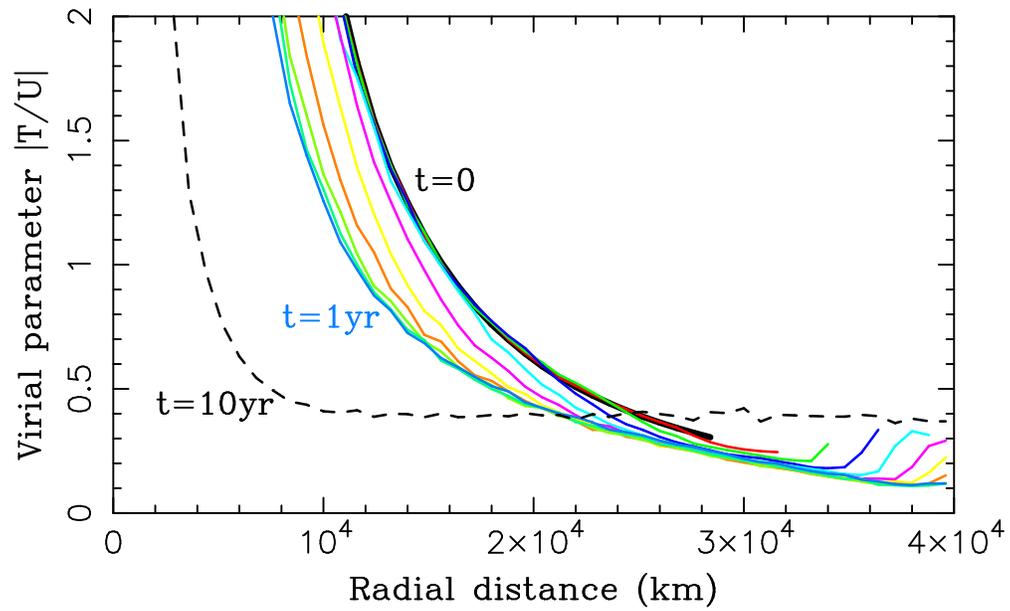}
\caption{The ratio of kinetic ($T$) and potential ($U$) energies as a function of radial distance. The line styles and 
parameters used here are the same as in Fig. \ref{profile}. The potential energy was computed as 
$U=-GM({\cal R})/{\cal R}$, where $M({\cal R})$ is the mass of particles inside a sphere of radius ${\cal R}$.}
\label{virial}
\end{figure}

Whereas this kind of behavior of radial profiles and velocity distributions is not predicted by the simple arguments in 
Section 2, the collapse 
timescale is roughly that of the free fall given by Eq. (\ref{ff2}). Recall that the simulated particles 
were initially given random velocities $v_{\rm rand}=0.8$ m s$^{-1}$. The velocities increase in the inner 
region during the free fall stage (Fig. \ref{profile}b). After that, collisions act to virialize 
the system\footnote{The escape speed from individual RPs in our tests is only $v_{\rm esc}=4.5$ cm s$^{-1}$, i.e., 
$\sim$20 times smaller than the random speed. The effect of gravitational scattering during particle 
encounters is therefore negligible. Note that even with radius inflation more massive SPs exaggerate the 
effect of gravitational scattering. The standard radius inflation by $\sqrt{n}$ gives an escape speed 
that increases as $n^{1/4}$ for SPs that are $n$ times as massive as their RPs. This is not a problem 
for the tests presented here because $v_{\rm esc}<v_{\rm rand}$ and the gravitational collapse timescale is 
relatively short. Gravitational stirring of SPs may become a problem in applications that will require large
values of $n$. In such cases, it would be desirable to soften the gravitational force/potential for 
neighbor SPs, but this will not completely solve the problem, because of the substantial contribution to 
stirring from distant SP encounters. This is an important limitation of the superparticle method.}  
and the radial velocity profiles become flatter (dashed line in 
Fig. \ref{profile}b). Eventually, the whole cloud, except for its crowded inner region, evolves 
toward the virial equilibrium at each radius (Fig. \ref{virial}). The slightly sub-virial conditions in 
the outer region ($T/|U| \simeq 0.4$ for $t=10$ yr; Fig. \ref{virial}) can be explained if particles are 
preferentially located near their orbital apocenters.  
 
Additional simulations were performed with the SP method described in Section 3. To start with, we used $10^3$ 
SPs each representing $n=10^3$ RPs. The simulations were done in the elastic regime and were thus not meant to 
test Eq. (\ref{eq:sigij}). The total energy change and departure from the virial equilibrium were both found 
to be slightly larger than in the fiducial case, probably due to the increased graininess of particle 
interactions. We also monitored the behavior of the SP system to see whether it reproduced Figs. \ref{profile} 
and \ref{virial}. We found that, indeed, the behavior was very similar and the radial profiles, random 
velocities and energies closely followed the trends described above. More information on these comparisons 
is provided below. The results with $10^4$ and $10^5$ SPs were found to be nearly identical to those obtained 
with $10^6$ RPs. The $10^2$ SP case, however, already significantly deviated from the reference case in that 
it showed a far weaker particle concentration near the center. This happened because SPs were strongly inflated 
and reached maximum packing in the center well before the collapse was completed.    

\subsection{Inelastic collisions}

The main goal of the tests reported here was to verify whether the SP method described in Section 3 correctly 
reproduces the energy loss. To this end we adopted $C_{\rm R}<1$ and varied the algorithm by which RPs were 
assigned to their parent SPs. We considered cases with equal and unequal sizes of RPs. For example, we used 
a power-law size distribution with $q=3$ and distributed RPs within a factor 4 in size around $r\simeq35$ km 
(i.e., the value used in the equal-size case). The most massive RPs  thus had 64 times the mass of the least massive
RPs. The SP method was tested for power-law distributions with $Q=1$, 2 and 3 (Sect 3.1).
  
\begin{figure}
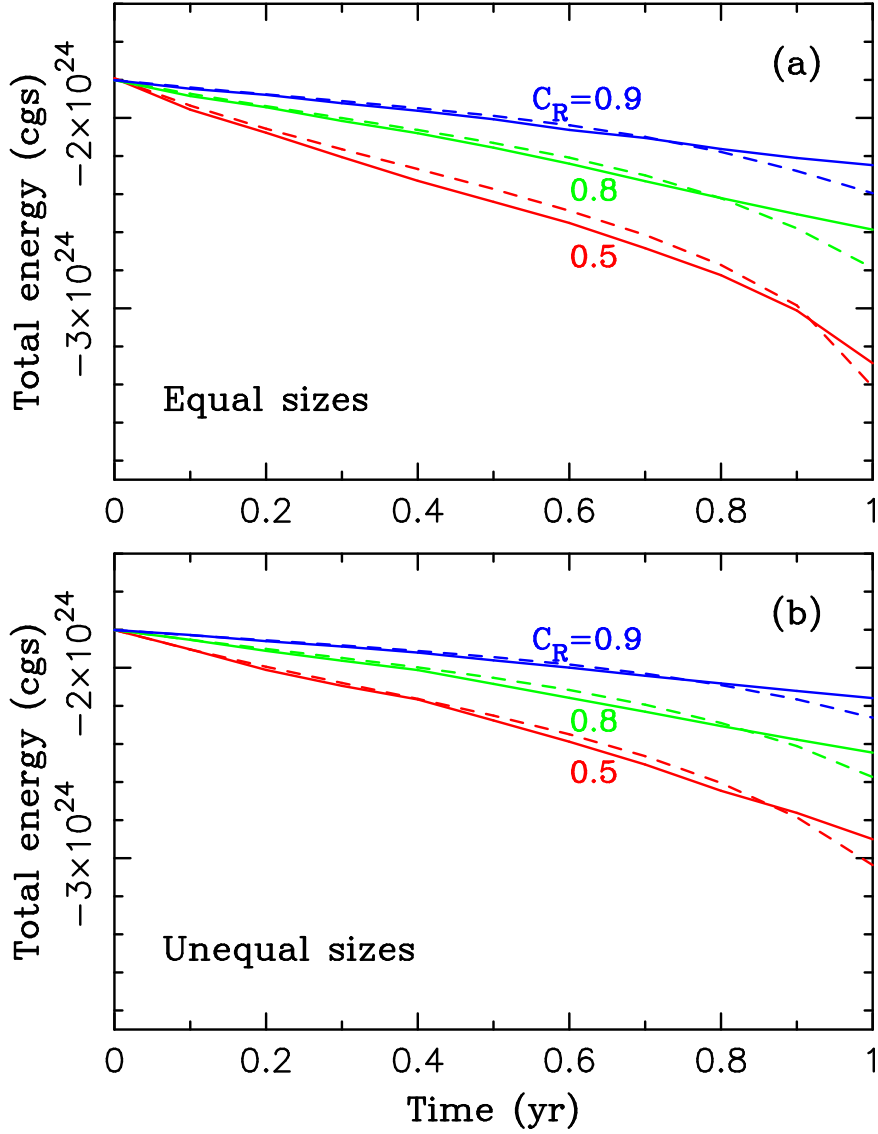

\epsscale{0.7}
\plotone{fig3a.eps}\\[3.mm]
\plotone{fig3b.eps}
\caption{The change of total energy for a case with equal-size particles (panel a)
and unequal-size particles (panel b). Different lines correspond to $C_{\rm R}=0.5$ (red), 0.8 (green) 
and 0.9 (blue). In each case, the solid line shows the result obtained by simulating $10^6$ particles. 
The dashed lines show the results obtained with $10^3$ SPs. For the 
unequal-size case, we used $q=2$ and $Q=1$ (Section 3.1). The results with $Q=2$ and $Q=3$ are nearly identical to the ones 
shown here for $Q=1$. We were unable to determine the energy loss beyond roughly one free fall timescale, because 
particles became crowded in the center and the simulations stalled just past $t=1$ yr due to an excessive 
number of collisions.}
\label{energy2}
\end{figure}

We experimented with $C_{\rm R}=0.5$, 0.8 and 0.9. The total energy loss from Eqs. (\ref{de}) and (\ref{dde})
is expected to be 1.5, 0.7 and 0.4 ($\times 10^{24}$ in cgs units per year; Eq. (\ref{energy}) does not 
apply, as we discussed in Sect.\ \ref{sec:freefall}, because the collapse happens on the free fall timescale). For 
comparison, the energy loss in a yr-long interval of the simulation with $10^6$ RPs was 1.6, 1 and 0.6
($\times 10^{24}$ in cgs units), respectively (Fig. \ref{energy2}). This is reasonable. An exact match is 
not expected because the equations describe an average behavior that doesn't account for spatial 
variations (or the time evolution of these variations) in the cloud. 
%(AY: I think more explicit but you decide)}  %the distribution of particles changes in the 
%simulaton and  influences the energy loss.

\begin{figure}
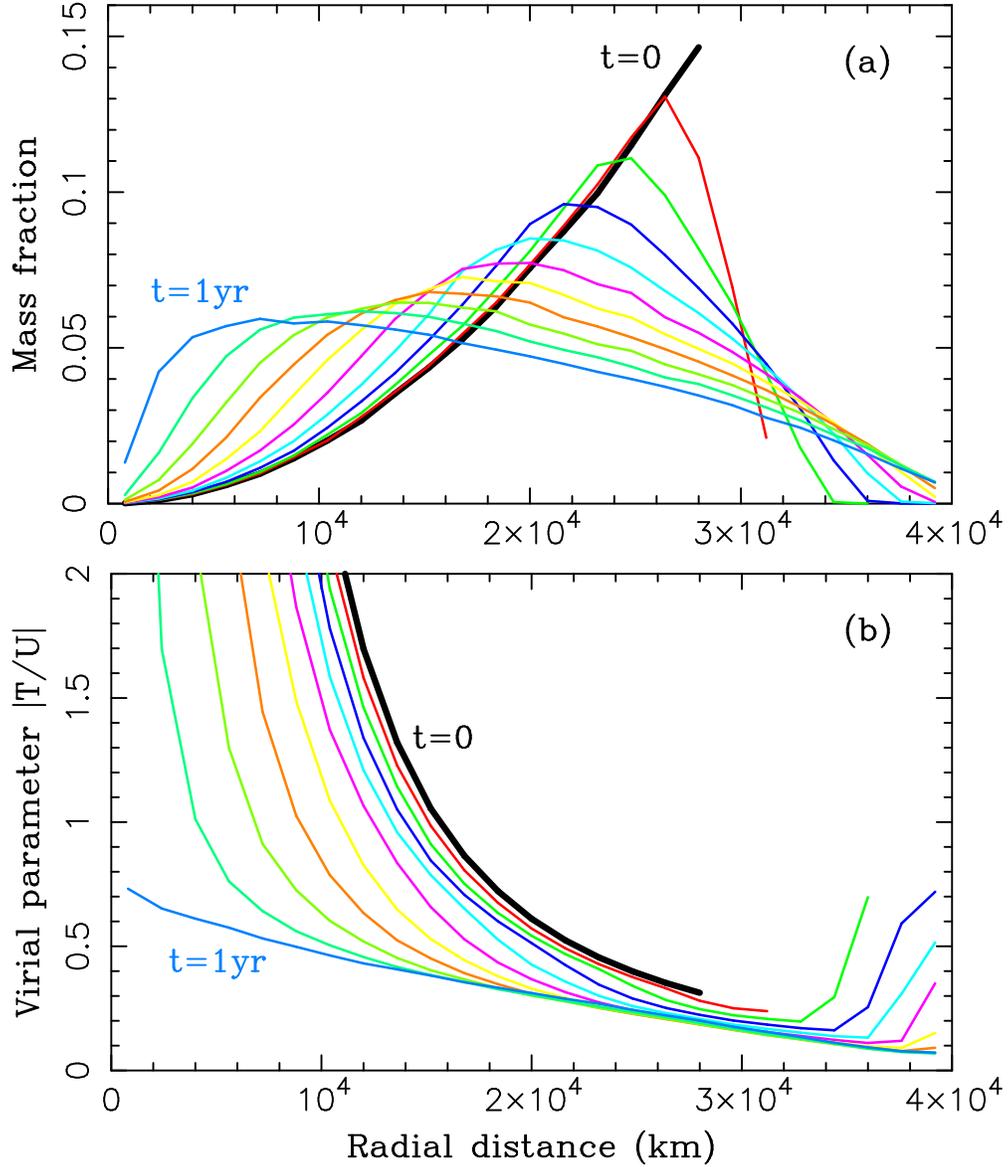

\epsscale{0.8}
\plotone{fig4a.eps}\\[2.mm]
\plotone{fig4b.eps}
%\plotone{meanv2.eps}
% gr.profile.f and gr.meanvel.f in COLLAPSE
\caption{The mass (a) and energy (b) profiles of a self-gravitating collisional cloud of particles. The line 
styles and simulation parameters used here are the same as in Fig. \ref{profile}, except in this case $C_{\rm R}=0.5$. The random velocities
are damped by inelastic collisions resulting in a stronger particle concentration in the inner region
(compare panel (a) with Fig. \ref{profile}a). The simulation stalls just after $t=1$ yr due to an excessive number
of collisions in the inner region.}
\label{profile2}
\end{figure}

The superparticle method with $10^3$ SPs replicated the energy loss reasonably well (Fig.~\ref{energy2}). 
The small difference between the solid and dashed lines in Fig.\ \ref{energy2} is a consequence of a relatively 
small number of SPs. The results with $10^4$ and $10^5$ SPs more closely replicated the reference case.
The energy loss was more random and a factor of $\sim$2 too small overall with only 100 SPs. We 
therefore focus on the $10^3$ SP case in the following tests, but comment on the results with 
other sampling rates as well. The radial mass profiles obtained in the case with $C_{\rm R}$=0.5 
(maximum dissipation tested here; Fig. \ref{profile2}a) are similar to those obtained in the elastic case 
(Fig. \ref{profile}a) except that the particles are even more crowded toward the center, and this trend becomes 
stronger with time. This is a consequence of the energy loss in inelastic collisions that damp random velocities 
and facilitate collapse. Eventually, as the number of collisions increased beyond reasonable limits in the 
central region, the code was unable to make any progress. We stopped all simulations at $t=1$ yr. 

\begin{figure}
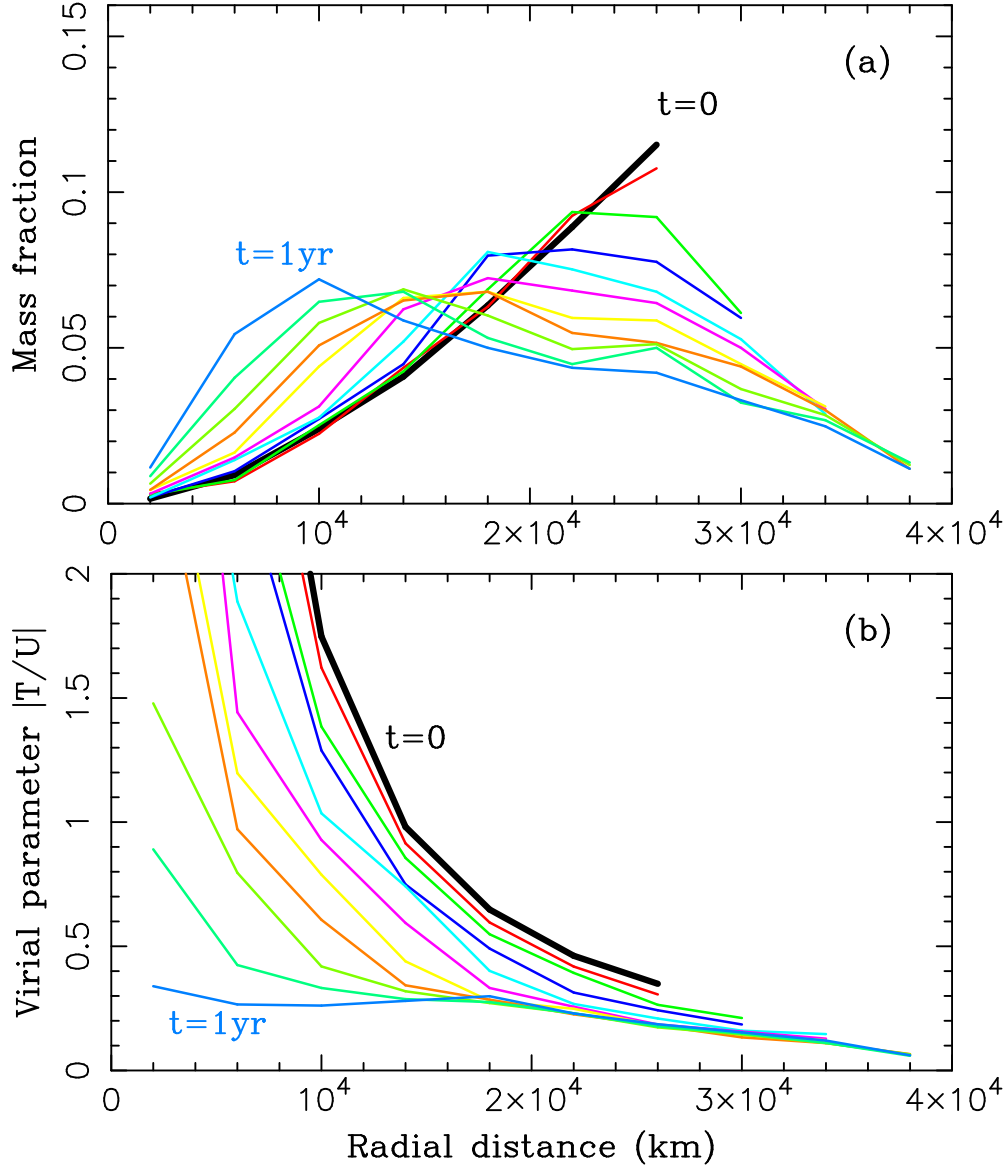

\epsscale{0.8}
\plotone{fig5a.eps}\\[2.mm]
\plotone{fig5b.eps}
\caption{The mass (a) and energy (b) profiles of a self-gravitating collisional cloud. The 
simulation setup is the same as in Fig. \ref{profile2}, except that here we used $10^3$ SPs
each representing $n=10^3$ RPs. The radial bin size was increased to compensate for poor resolution 
(10 bins here vs. 25 bins in previous figures). Panel (a) can be compared to Fig. \ref{profile2}a 
and panel (b) to Fig. \ref{profile2}b.}
\label{profile3}
\end{figure}

The virial parameter $T/|U|$ is shown in Fig. \ref{profile2}b. The $T/|U|$ ratio is generally smaller than that 
in Fig. \ref{virial} at least partly because here the kinetic energy decays by dissipative collisions. This becomes 
especially clear near the center as time approaches the free fall timescale. All these trends are well reproduced with 
the SP method (Fig. \ref{profile3}). As the SPs occupy a larger volume, however, they are not allowed to 
concentrate as strongly toward the center as in the reference case. This leads to the mass distribution 
that peaks at somewhat larger radial distance (compare Fig. \ref{profile2}a and Fig. \ref{profile3}a). 
This ``packing'' problem becomes less (more) of an issue with $>10^3$ SPs ($<10^3$ SPs). In addition, the radial 
profiles become very noisy with only $10^2$ SPs, suggesting that this sampling rate is not adequate to 
capture the statistical behavior of the real system. 

We verified that the total angular momentum $L$ is conserved by adding solid initial rotation 
of the cloud around the $z$ axis. At the cloud's outer radius of $\simeq$30,000 km, the circular speed is 
$v_{\rm circ}=\sqrt{G M / R} \simeq 1.1$ m s$^{-1}$. We thus added angular speed $\Omega = f_\Omega v_{\rm circ}/R$, 
where $f_\Omega \geq 0$ is a scaling factor. Specifically, for the tests in the following section, we used 
$f_\Omega=0.5$ and $\Omega \simeq 1.7\times 10^{-8}$~s$^{-1}$. This choice was motivated by rotation inferred
for self-gravitating pebble clouds from our SI simulations (e.g., Nesvorn\'y et al. 2019, Li et al. 2019),
but note that the initial structure of SI clouds is generally complex and cannot be idealized by solid rotation
of homogeneous spheres. Here we used the idealized case to test the SP method. More realistic initial 
conditions will be investigated elsewhere. We found that the angular momentum was very well conserved in 
both the reference case (relative change $\delta L/L \sim 3 \times 10^{-6}$) and when the 
SP method was used (e.g., $\delta L/L \sim 5 \times 10^{-5}$ for the case with $10^3$ SPs). 

\subsection{Mergers}

Here we tested the merger algorithm described in Section 3.3. The general setup of the simulations was the same 
as the one used above. We tested cases with ($f_\Omega=0.5$) and without ($f_\Omega=0$) cloud rotation. 
The initial rotation provides centrifugal support to the cloud and reduces problems with particle crowding.
We found that the case with $f_\Omega=0.5$ is more suitable for comparisons than $f_\Omega=0$ because the results 
are less sensitive to various integration details (e.g., the seed used to generate the initial conditions).
The results for $f_\Omega=0$ are reported below for completeness. In {\tt PKDGRAV}, particle mergers are 
applied if the collision speed between particles $v_{\rm col} < f_{\rm esc} v_{\rm esc}$, where $v_{\rm esc}$ is the 
escape speed and $f_{\rm esc}<1$ is a free parameter. Collisions with $v_{\rm col} > f_{\rm esc} v_{\rm esc}$ result
is particle bounces.  

\begin{figure}
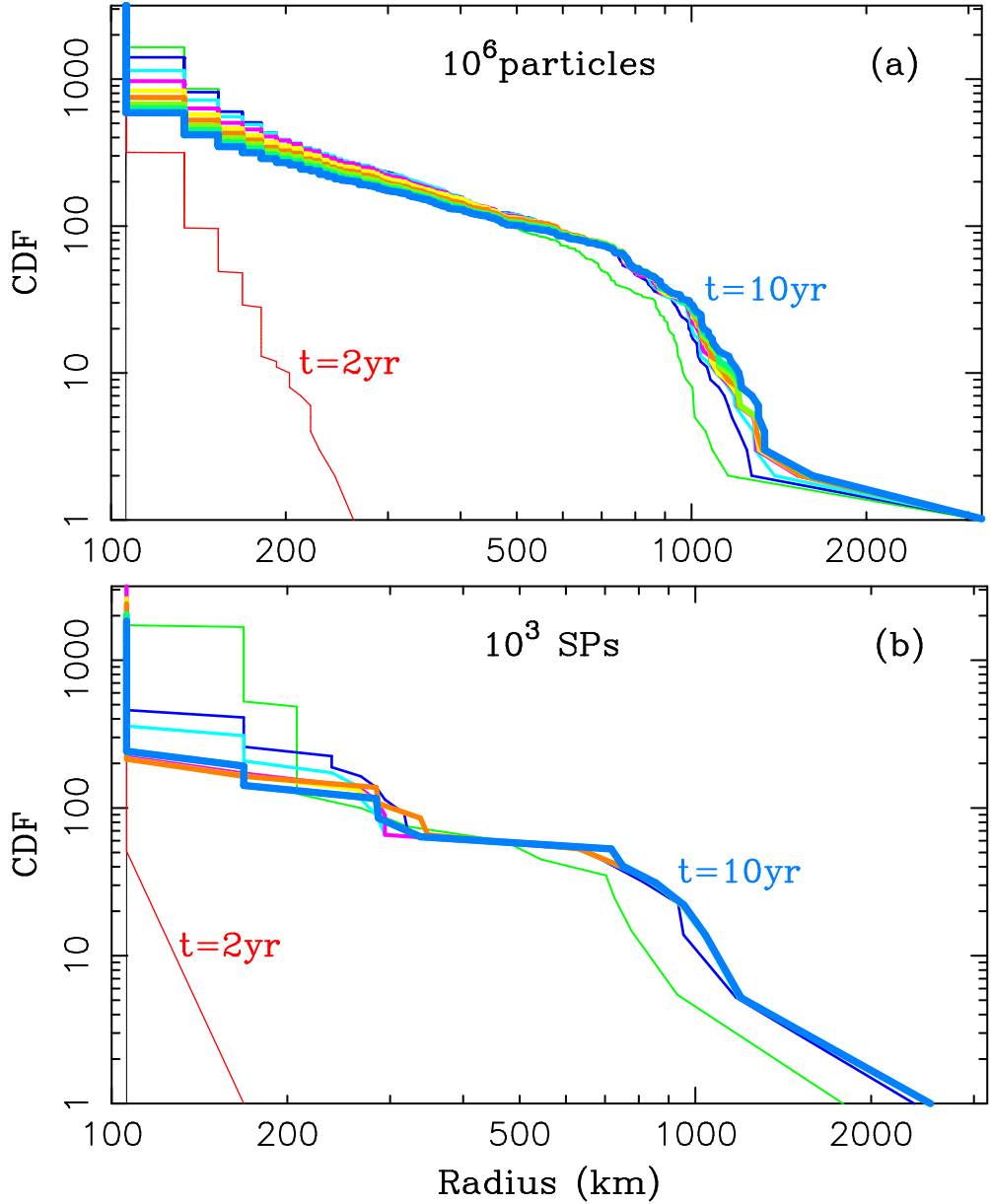

\epsscale{0.8}
\plotone{fig6a.eps}\\[2.mm]
\plotone{fig6b.eps}
\caption{The cumulative size distributions of particles for the case with $C_{\rm R}=0.5$ and $f_{\Omega}=0$.
Mergers between particles occur in {\tt PKDGRAV} when the collision speed is less then 0.1 of the particle 
escape speed. The size distribution is shown for $0<t<10$ yr in 1~yr increments (from thin black to 
thick blue lines). The reference case with $10^6$ particles is shown in panel (a). The case with $10^3$ SPs is shown in panel (b).}
\label{merge2}
\end{figure}

Figure \ref{merge2} shows the size distributions obtained in the simulations with $10^6$ RPs and $10^3$ SPs. 
In both cases, we used $v_{\rm rand}=0.8$ m s$^{-1}$, $C_{\rm R}=0.5$, $f_\Omega=0$ and $f_{\rm esc}=0.1$. 
In the reference case, we 
simply plot the size distribution of RPs. In the SP simulation, the size distribution is constructed from 
the number and sizes of particles represented by each SP. We found that the SP algorithm with $10^3$ SPs can 
reproduce the particle growth quite well but the results had a rather large statistical variability with minor 
changes of the input (e.g., when different seeds were used to generate initial conditions). We interpret 
this as a consequence of stochastic interaction of large bodies that grow in the center of the collapsing cloud.
The stochasticity of the results diminishes if $>10^3$ SPs are used. For this reason, in the rest of this 
section, we discuss the results obtained with $f_\Omega=0.5$, as rotation inhibits collapse to the center of mass. 
For reference, the radii of the largest bodies shown in Fig. \ref{merge2} are $\simeq$3200 km (panel a) and 
$\simeq$2500 km (panel b). 

\begin{figure}
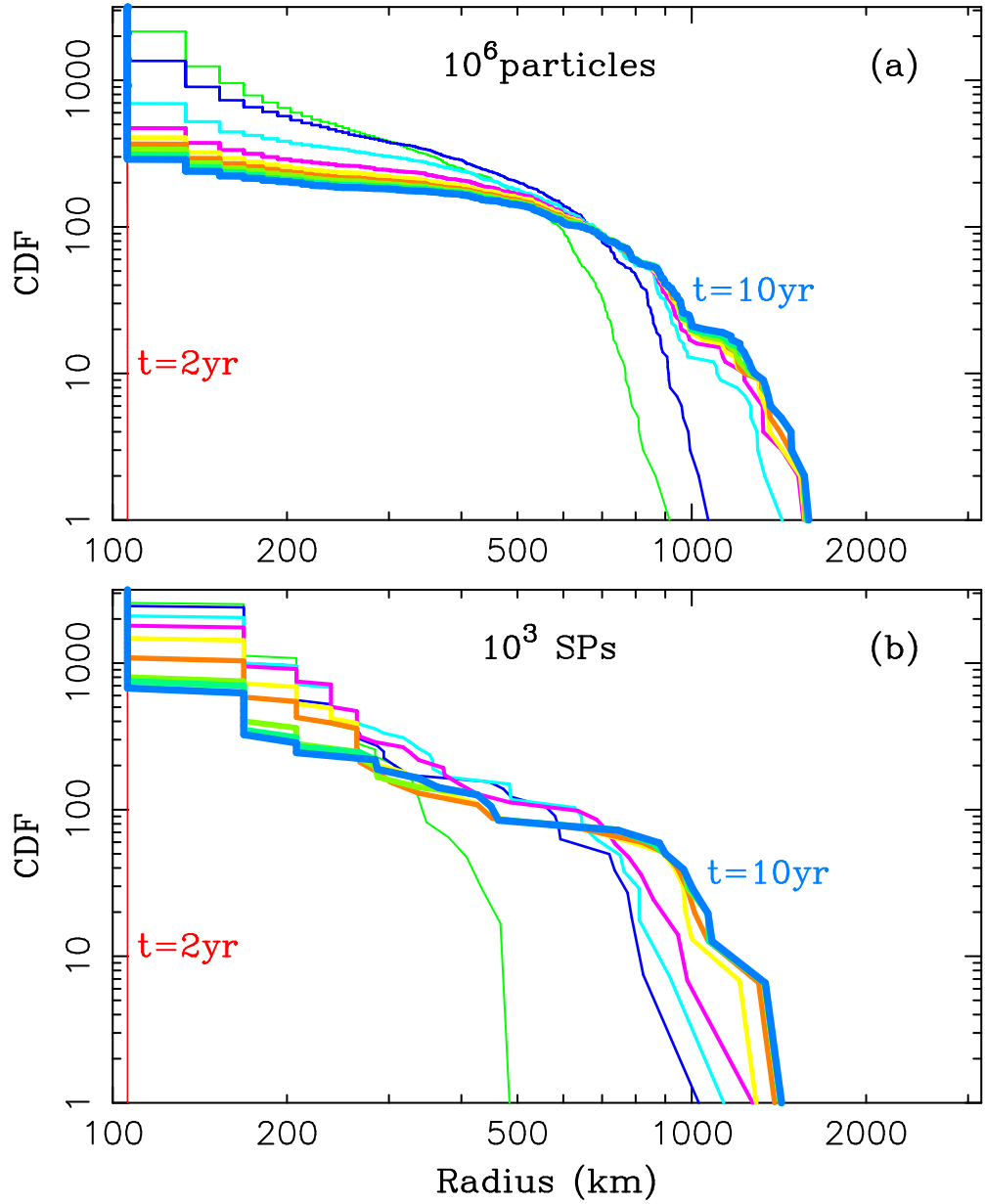

\epsscale{0.8}
\plotone{fig7a.eps}\\[2.mm]
\plotone{fig7b.eps}
\caption{The cumulative size distributions of particles for the case with $C_{\rm R}=0.5$ and $f_{\Omega}=0.5$.
See caption of Fig. \ref{merge2} for additional information.}
\label{merge1}
\end{figure}

\begin{figure}
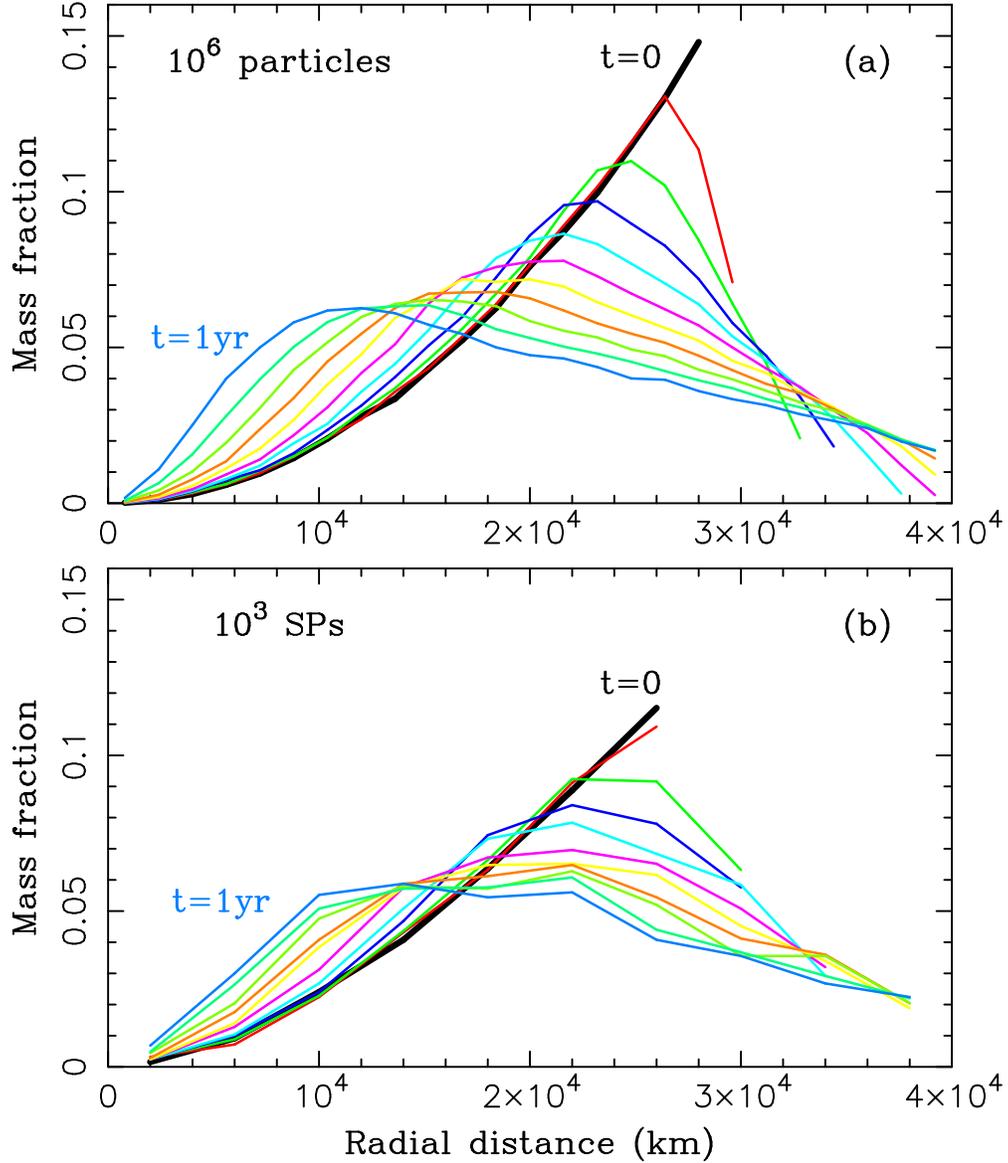

\epsscale{0.8}
\plotone{fig8a.eps}\\[2.mm]
\plotone{fig8b.eps}
\caption{The radial profiles for the case in Fig. \ref{merge1}. The profiles are shown for $0<t<1$~yr 
in 0.1 yr increments. The profiles become noisy for $t>1$ yr (not shown here) when large bodies grow 
in the cloud. Panels (a) and (b) show the cases with $10^6$ particles and $10^3$ SPs, respectively.}
\label{merge_prof1}
\end{figure}

The case with $f_\Omega=0.5$ does not produce bodies that are as large (Fig. \ref{merge1}). The largest objects have 
the radii $\simeq$1600 km (panel a; $10^6$ RPs) and $\simeq$1400 km (panel b; $10^3$ SPs). Both simulations ended with 
a rounded cumulative distribution function (CDF) profile and dozens of bodies larger than 1000 km. For $f_\Omega=0.5$, 
the SP method somewhat overshoots the number of the smallest objects but the difference is not large (the case 
with $f_\Omega=0$ produced an opposite result; Fig. \ref{merge2}). We repeated the SP simulations with slightly 
changed initial conditions and found that the results were quite similar to those shown in Fig. \ref{merge1}b.
This means that the stochastic variations are not as large for $f_\Omega=0.5$ as they were for $f_\Omega=0$.
The radial mass distribution profiles are compared in Fig. \ref{merge_prof1}. The SP method with $10^3$ SPs 
reasonably well replicates the profiles obtained in the fiducial case. The results with $10^4$ and $10^5$ SPs 
reproduce the fiducial radial profiles and size distributions even more closely.   

\begin{figure}
\epsscale{0.49}
\plotone{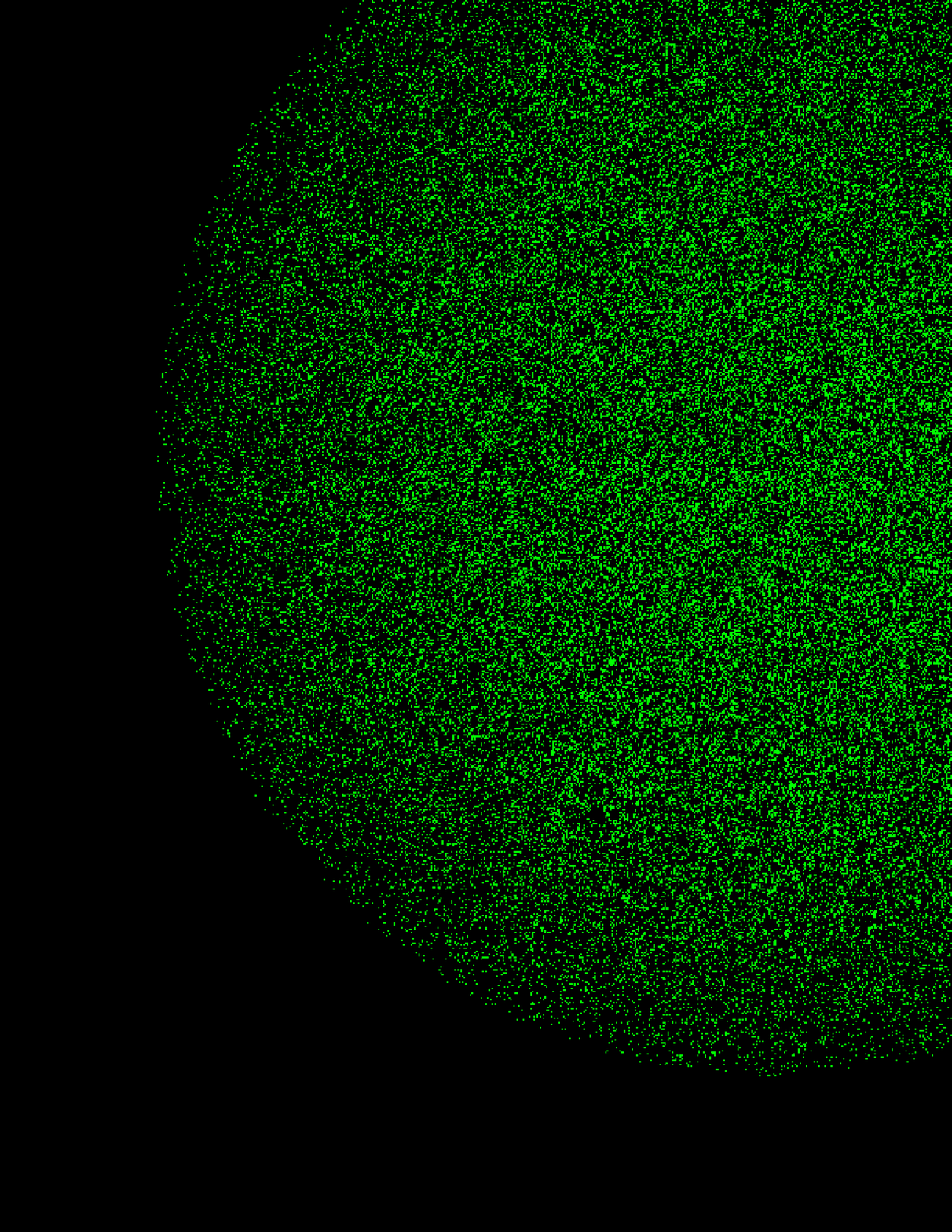}
\plotone{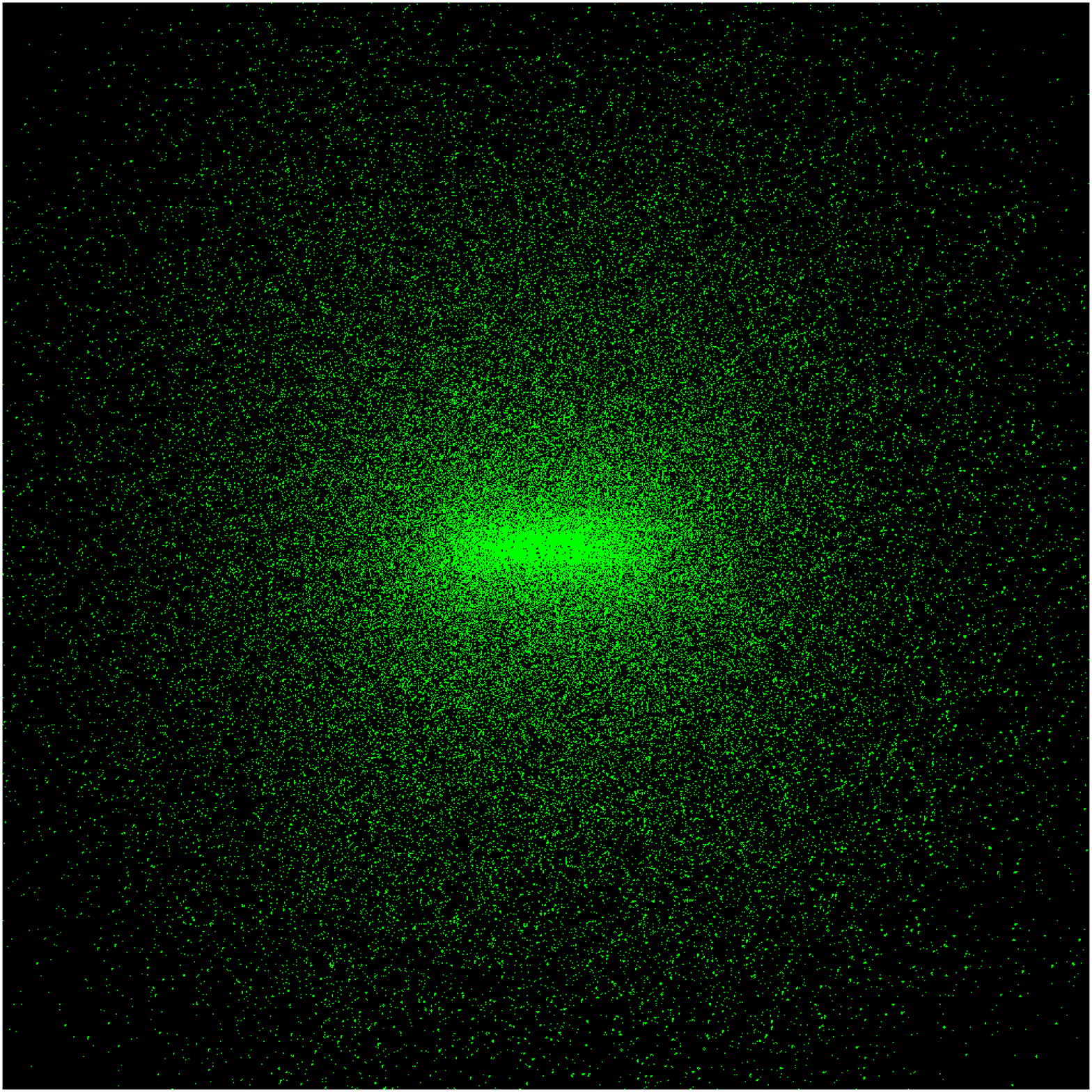}\\[1.mm]
\plotone{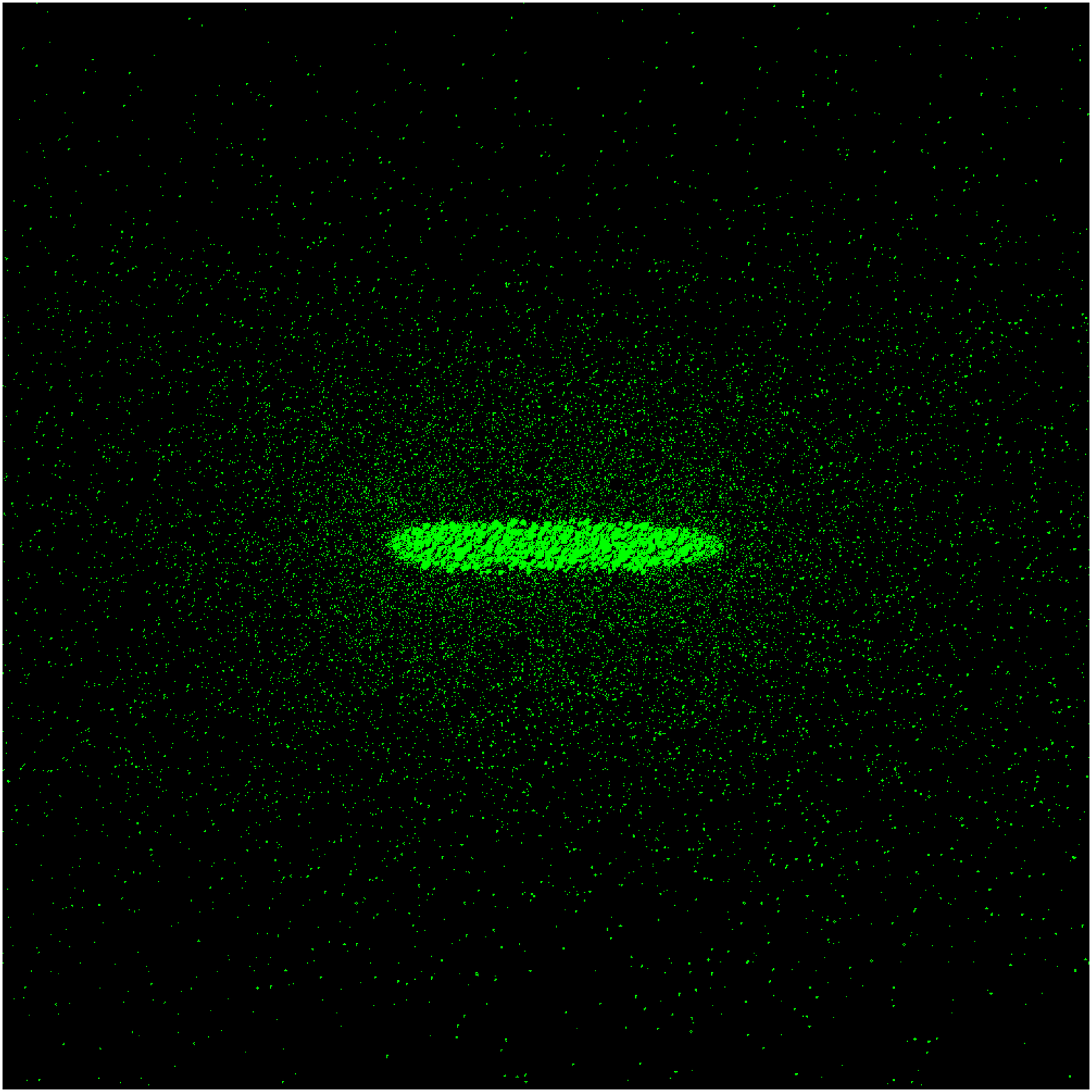}
\plotone{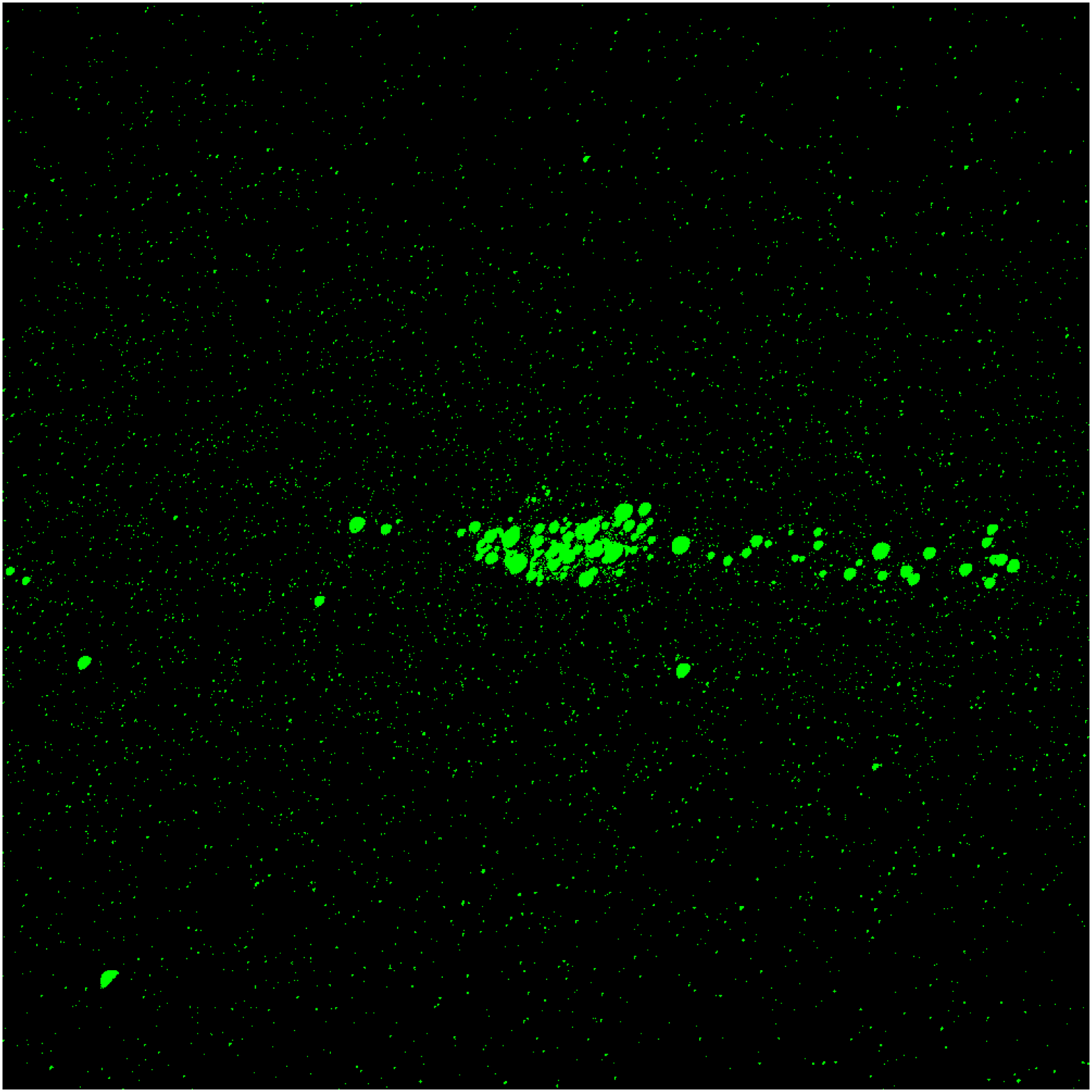}
\caption{Four snapshots showing particle distributions at $t=0$ (top left), 1.2 (top right), 3 (bottom left) 
and 7 years (bottom right). This is for the case with $C_{\rm R}=0.5$ and $f_\Omega=0.5$.}
\label{snaps}
\end{figure}

Figure \ref{snaps} shows the evolving structure of the particle cloud with $f_\Omega=0.5$. First, an agglomerate 
of particles formed in the center. The agglomerate had significant rotation and stretched along one direction 
to a $\simeq$7:1 axial ratio. Then, roughly at $t=4$ yr, the extremes detached to release the angular 
momentum whereas the middle part of the agglomerate remained connected. Several small binaries formed at 
this point. Eventually, when the simulation was extended to $t=100$ yr, a massive, equal-size binary formed 
in the center. The two components of the binary had radii $R_1=24.1$ km and $R_2=20.5$ km and a separation 
of $a_{\rm B}=4530$ km. Thus $a_{\rm B}/(R_1^3+R_2^3)^{1/3} \simeq 160$, which is quite typical for the equal-size 
binaries in the trans-Neptunian region (Grundy et al. 2019, Nesvorn\'y \& Vokrouhlick\'y 2019). 
The total mass of the binary object represents 18\% of the initial cloud's mass. This suggests that 
planetesimal masses estimated from the total cloud mass (e.g., Simon et al. 2017) could be overestimated by 
an order of magnitude.
 
%\subsection{Equipartition} Case {\tt Test2\_elast\_unequal}.

\section{Conclusions}

We developed a new method that will be useful for modeling statistically large collisional systems of 
particles. The method represents a significant computational speed-up by employing far fewer 
``superparticles'' than the number of particles in the real system. It reproduces the energy loss in 
dissipative collisions of real particles and can account for particle growth by mergers. We implemented 
the method in the {\tt PKDGRAV} code and tested it. We found that at least $\sim$$10^3$ superparticles 
need to be used to reproduce the results of our fiducial simulations with full resolution. The number
of superparticles needed in real-world applications will have to be determined by convergence studies 
tailored to those applications. As the superparticles need to be inflated to represent a much larger collisional 
cross-section of the real system, particle crowding may become a problem in some applications. In the 
gravitational collapse simulation performed here, this becomes apparent near the cloud's center, where the 
code has difficulties to account for a vast number of superparticle collisions. This problem can be mitigated 
by using a larger number of (less-inflated) superparticles, merging superparticles and/or letting {\tt PKDGRAV} 
form rigid particle aggregates.
        
\acknowledgements
The work of DN was funded by the NASA EW and XRP programs. ANY acknowledges support from 
NASA Astrophysics Theory Grant NNX17AK59G and from NSF grant AST-1616929. RM acknowledges the 
support of the Swiss National Science Foundation (SNSF) through the grant P2BEP2\_184482. 
We thank the anonymous reviewer for helpful suggestions.   

{\bf Author's contributions:} DN motivated this study, performed and analyzed all tests with {\tt PKDGRAV},
and prepared the article for publication. ANY derived Eq. (\ref{eq:sigij}) and wrote Section 3. RM 
helped to develop tests that are reported in Section 4. DCR provided support related to the {\tt PKDGRAV} 
code. All authors contributed to the interpretation of the results and writing this paper.

%\begin{figure}
%\epsscale{0.8}
%\plotone{virial2.eps}
%% gr.virialp.f 
%\caption{The ratio of kinetic ($T$) and potential ($U$) energies as a function of radius. The line styles and 
%parameters used here are the same as in Fig. \ref{profile}, except for $C_R=0.5$. Inelastic collisions result
%in kinetic energy loss and progressively lower $|T/U|$ values than in Fig. \ref{virial}.}
%\label{virial2}
%\end{figure}

\end{document}